\documentclass[letterpaper]{JHEP3}
\usepackage{cite}
\newcommand{\fft}[2]{{\frac{#1}{#2}}}
\newcommand{\ft}[2]{{\textstyle\frac{#1}{#2}}}
\newcommand{\Tr}{{\rm Tr\,}}
\title{Supersymmetry and the AdS Higgs Phenomenon}

\author{Benjamin A. Burrington and James T. Liu\\
Michigan Center for Theoretical Physics\\
Randall Laboratory of Physics, The University of Michigan\\
Ann Arbor, MI 48109-1120\\
E-main: \email{bburring@umich.edu}, \email{jimliu@umich.edu}}

\preprint{MCTP-03-48\\
\hepth{0311205}}

\abstract{We examine the AdS Higgs phenomenon for spin-1 fields, and
demonstrate that graviphotons pick up a dynamically generated mass in
AdS$_4$, once matter boundary conditions are relaxed.  We perform an
explicit one-loop calculation of the graviphoton mass, and compare this
result with the mass generated for the graviton in AdS.  In this manner,
we obtain a condition for unbroken supersymmetry.  With this condition,
we examine both ${\cal N}=2$ and ${\cal N}=4$ gauged supergravities coupled
to matter multiplets, and find that for both cases the ratio between
dynamically generated graviton and graviphoton masses is consistent with
unbroken supersymmetry.}


\begin{document} 


\section{Introduction}

It has long been known that field theories in curved spacetime may exhibit
interesting behavior without corresponding flat space analogs.  While
field theories in maximally symmetric spaces have already been well explored,
recent developments in (A)dS/CFT and cosmology have led to renewed
interest in field theories in de Sitter and anti-de Sitter spaces.  For
the latter, it is noteworthy that AdS space is not globally hyperbolic
because of its timelike boundary, which is easily reached in finite time
by null geodesics.  Traditionally, the boundary of AdS has been dealt with
by imposing reflecting boundary conditions, which is natural from the point
of view of information flow \cite{Avis:1977yn}.  However, investigations of
the holographic dual to the Karch-Randall model \cite{Karch:2000ct} has
led to the realization that the presence of multiple holographic domains
\cite{Porrati:2001gx,Bousso:2001cf} naturally leads to transparent boundary
conditions for the AdS field theory on the brane.

Thereafter, it was demonstrated by Porrati that a simple field theory in
AdS with transparent boundary conditions coupled to Einstein gravity
(with conventional reflecting boundary conditions for the graviton), leads
to a dynamical generation of graviton mass \cite{Porrati:2001db}.
This mechanism has been denoted the AdS Higgs phenomenon
\cite{Porrati:2001db,Porrati:2003sa}, as the graviton gets mass by
eating a composite Goldstone vector, which is a kinematically bound state
of the field theory particles.  While the one-loop computation of
\cite{Porrati:2001db} was for a single conformally coupled scalar with
generalized boundary conditions, it was subsequently demonstrated in
\cite{Duff:2002ab} that matter fields of spins $0$, $1/2$ and $1$ all
contribute towards the graviton mass.  In particular, viewing the holographic
dual to the Karch-Randall model as a ${\cal N}=4$ super-Yang-Mills CFT
with $U(N)$ gauge group, it was shown in \cite{Duff:2002ab} that the
one-loop AdS Higgs computation correctly reproduces the result for the
Karch-Randall graviton mass.

An important fact about the AdS Higgs phenomenon is that, while a
mass is generated for the graviton, general covariance remains unbroken,
and the gravitational Ward identities remain satisfied 
\cite{Porrati:2001db,Porrati:2003sa}.  Similarly, one may imagine from
a supersymmetric context that if the graviton were to get a mass, its
superpartners ought to become massive as well.  In this sense, supersymmetry
may remain unbroken, even with a dynamically generated gravitino mass, as in
the case of general covariance and graviton mass.  We will show
that this is indeed
what happens for supergravity coupled to a CFT, where the CFT fields are
given unusual boundary conditions.  This result is at least {\it consistent}
with a supersymmetric realization of the Karch-Randall braneworld, with the
entire localized supergravity multiplet originating from the quasi-zero-mode
part of the Kaluza-Klein tower.

The AdS Higgs mechanism may be concisely stated in terms of $SO(2,3)$
representation theory.  For irreducible representations labeled by
$D(E_0,s)$, massless representations in AdS$_4$ (as defined by propagation
of a reduced number of states) generically corresponds to $D(E_0=s+1,s)$, so
that {\it e.g.}~a massless graviton is given by $D(3,2)$ and a massless
vector is given by $D(2,1)$.  In this case, the AdS Higgs mechanism
corresponds to the decomposition of a massive representation $D(E_0>s+1,s)$
in the massless limit
\begin{equation}
\lim_{\epsilon\to0}D(s+1+\epsilon,s)=D(s+1,s)+D(s+2,s-1)\qquad(s\ge1).
\label{eq:adshlim}
\end{equation}
In particular, the graviton gets a mass by eating a massive $D(4,1)$ vector,
while a graviphoton (or ordinary photon, for that matter) gets a mass
by eating a minimally coupled $D(3,0)$ scalar.  The latter case of course
corresponds to the ordinary Higgs mechanism, whether in AdS or in flat space.

In this paper, we show that the graviphoton (in either ${\cal N}=2$ or
${\cal N}=4$ theories) indeed picks up a mass through the AdS Higgs
phenomenon, and furthermore that its mass is related to the graviton mass
in precisely the ratio demanded by the preservation of supersymmetry
(for the massive spin-$2$ AdS multiplet).  Of course, we expect the
gravitino to become massive in just the same manner, although we have not
performed an explicit check.  This AdS Higgs phenomenon is in fact quite
general, regardless of spin, and it seems fair to say that mass generation
in AdS is a generic phenomenon any time boundary conditions are relaxed.

We begin in section~2 with a general discussion of the AdS Higgs phenomenon
as it applies to mass generation for vector fields.  The method for
identifying the photon mass parallels that for the graviton mass worked out
in \cite{Porrati:2001db,Duff:2002ab}.  In section~3, we perform the actual
one-loop calculation for both scalars and spin-1/2 fermions running through
the loop.  This is sufficient for examining graviphoton masses in ${\cal N}=2$
and ${\cal N}=4$ supergravities, which we do in section~4.  Finally, we
conclude in section~5 by highlighting some of the main features of the
AdS Higgs phenomenon.


\section{The spin-1 AdS Higgs phenomenon}

Although masslessness of the photon (or graviton) is traditionally associated
with gauge invariance, the actual connection is rather more subtle.  This
was emphasized in \cite{Porrati:2001db} in the context of the AdS Higgs
phenomenon.  In fact, the kinematics of mass generation for a spin-1
vector is simpler than that of the graviton, and we review this here.

Ordinary field theory in AdS, or any curved space for that matter, is
somewhat more involved than in a flat Minkowski background.  In particular,
lack of translational symmetry precludes a straightforward momentum space
treatment.  Nevertheless, we may proceed with a coordinate space analysis.
If $\Sigma^\mu{}_\nu(x,y)$ denotes the vector boson self-energy, then
gauge invariance ensures the transversality of $\Sigma^\mu{}_\nu$, so that
it may be written in the form
\begin{equation}
\label{fself}
  \Sigma^{\mu}{}_{\nu}=\beta(\Delta)\Pi^{\mu}{}_{\nu},
\end{equation}
where $\beta(\Delta)$ is a scalar function (containing the dynamics, and
undetermined by transversality) and $\Pi^\mu{}_\nu$ is the transverse
projection
\begin{equation}
\label{trans}
  \Pi^{\mu}{}_{\nu}=g^{\mu}{}_{\nu}+\nabla^{\mu}\frac{1}{\Delta} 
  \nabla_{\nu}.
\end{equation}
Here, $\Delta$ denotes the Lichnerowicz operator which commutes with
covariant differentiation, $[\Delta,\nabla_\mu]T_{\{\mu_i\}}=0$.  In
particular, 
$\Delta^{(0)}\phi=-\nabla^2\phi$ and $\Delta^{(1)}A_{\mu}=-\nabla^2 A_{\mu}
+R_{\mu\nu}A^{\nu}$ when acting on scalars and vectors, respectively.

Working in Landau gauge, the bare vector propagator takes the form
\begin{equation}
  D^{\mu}{}_{\nu}=\frac{\Pi^{\mu}{}_\nu}{\Delta}.
\end{equation}
Since $\Pi^\mu{}_\nu$ is a projection, the full propagator is easily
resummed, yielding
\begin{equation}
  \tilde{D}^\mu{}_\nu=\frac{\Pi^\mu{}_\nu}{\Delta-\beta(\Delta)},
\end{equation}
which is physical when evaluated between conserved currents.  Any potential
mass may then be read off by the shift in the pole of the propagator.
Therefore it is the constant piece in the expansion $\beta(\Delta)
=-M^2+{\cal O}(\Delta)$ that yields a photon mass.  This argument is simply
the curved space analog of the standard textbook one, where
$\Sigma^\mu{}_\nu=\beta(p^2)(\delta^\mu{}_\nu-p^\mu p_\nu/p^2)$ results
in a resummed expression $\tilde D=\delta^\mu{}_\nu/(p^2+\beta(p^2))+\cdots$.
Photon mass is then generated if $\beta(p^2)=-M^2+{\cal O}(p^2)$.  It is
clear from the flat space context that a shift in the photon pole arises
from a {\it non-local} self-energy expression.  Likewise, this continues
to be the case in curved space.  This potential non-locality is the basis
for identifying the actual value of the dynamically generated photon (or
graviton) mass \cite{Porrati:2001db,Duff:2002ab}.

While representation theory in AdS allows the possibility of an AdS Higgs
phenomenon as indicated in (\ref{eq:adshlim}), we must perform an actual
loop computation to see that the generated mass is non-vanishing.  This
computation follows the techniques developed in
\cite{Porrati:2001db,Duff:2002ab}.  The method is essentially to compute
the one-loop self energy, extract its non-local behavior, and then to
identify the actual mass from the non-local data.

Let us actually consider the last point first, namely identifying the
dynamically generated mass from the self-energy.
We find it easiest to continue in homogeneous coordinates by embedding
AdS$_4$ into $R^5$. In this case, AdS$_4$ is the restriction to the
hyperboloid in $R^5$ given by $X^{M}X_M=-L^2$, where $R^5$ has metric
$\eta_{MN}={\rm diag}(-,+,+,+,-)$.  We use $X^M$, $Y^N$ ($M,N=0,\ldots,4$)
to denote homogeneous coordinates and $x^{\mu}$, $y^{\nu}$
($\mu,\nu=0,\ldots,3$) to denote intrinsic coordinates on AdS$_4$.  The
metric on AdS$_4$ is then the projection of the $R^5$ metric onto the
hyperboloid, $G^{MN}(X)=\eta^{MN}+X^MX^N/L^2$.  Note that $G^{MN}$ also
serves to project vector quantities onto AdS$_4$.

Scalar two-point functions $\phi(X,Y)$ in maximally symmetric spaces 
can only depend on the invariant interval between $X$ and $Y$.  This in
turn may be expressed in terms of the $R^5$ distance $|X-Y|^2$.  Since
$X^2=Y^2=-L^2$ when restricted to AdS$_4$, this indicates that $\phi(X,Y)$
may be written in terms of a single scalar invariant $Z\equiv X\cdot Y/L^2$.
Note that $|X-Y|^2/L^2=-2(Z+1)$, so that the short distance region corresponds
to taking $Z\to-1$.  In fact, the hyperboloid in $R^5$ has two branches, and
$Z$ has the values $-\infty\le Z\le-1$ on the `physical' AdS$_4$, while
$1\le Z\le\infty$ on the other branch.  Recalling that AdS space is conformal
to {\it half} of the Einstein static universe, the map $Z\to-Z$ ({\it i.e.}
$Y\to-Y$) may then be viewed as a map taking $Y$ to its image point on the
Einstein static universe.

Vector functions in AdS$_4$ may be treated similarly using homogeneous
coordinates.  However, we must keep in mind to project all free indices
onto AdS$_4$ using $G^{MN}$.  As a result, symmetric vector two-point
functions $\Sigma_{MN}(X,Y)$ may be written in terms of two invariant
bi-vectors, $\hat G_{MN}(X,Y)$ and $N_M(X)N_N(Y)$, where
\begin{eqnarray}
\hat{G}_{MN}(X,Y)&=&G_{ML}(X)\eta^{LP}G_{PN}(Y)
=\eta_{MN}+(X_MX_N+Y_MY_N+ZX_MY_N)/L^2,\nonumber\\
N_{M}(X)&=&\frac{Y_M+ZX_M}{L\sqrt{Z^2-1}}.
\end{eqnarray}
These expressions are the direct generalizations of the flat space
quantities $\eta_{\mu\nu}$ and $\hat n_\mu\hat n_\nu$ (where $\hat n$
is the unit vector pointing from $x^\mu$ to $y^\mu$).  For convenience,
we may decompose $\Sigma_{MN}(X,Y)$ as
\begin{equation}
\Sigma_{MN}=a(Z)\left(\hat{G}_{MN}-Z N_M N_N\right)
+ b(Z)N_M N_N,
\label{eq:smnab}
\end{equation}
where it is to be understood that the index $M$ refers to point $X$ and
the index $N$ refers to point $Y$.  In particular, arbitrary symmetric
bi-vectors are specified by two independent scalar quantities $a(Z)$ and
$b(Z)$.

\TABLE{
\begin{tabular}{c|cc}
&$a(Z)$&$b(Z)$\\
\hline
$T_3$&$0/Z^2+1/Z^4$   &$1/Z^3$\\
$T_4$&$-1/3Z^3+4/3Z^5$&$1/Z^4$\\
$T_5$&$-2/3Z^4+5/3Z^6$&$1/Z^5$\\
$T_6$&$-1/Z^5+2/Z^7$ &$1/Z^6$\\
\end{tabular}
\caption{Leading basis tensors for the long distance expansion of the
self-energy.}
\label{tbl:1}
}

Of course, transversality of the vector self-energy imposes a further
condition on $a(Z)$ and $b(Z)$.  To see this, we may take the covariant
divergence $\nabla^M\Sigma_{MN}=G^{ML}\partial_M\Sigma_{LN}$ of
(\ref{eq:smnab}) and demand that it vanishes.  This yields a differential
condition
\begin{equation}
a=\ft13(Z^2-1)\fft{db(Z)}{dZ}+Zb(Z),
\end{equation}
so that the transverse self-energy may be completely specified by a single
function $b(Z)$.  We note that a large distance expansion of $\Sigma_{MN}$
may be performed in terms of inverse powers of $Z$.  In this case, it is
convenient to introduce a set of asymptotic basis tensors $T_i$, so that
$\Sigma^{MN}=\sum_ib_iT_i^{MN}$ where $b_i$ are the constant coefficients
$b(Z)=\sum_ib_i/Z^i$.  The first few leading terms in the expansion are
given in Table~\ref{tbl:1}.

We now use these expressions to cast the transverse self-energy,
(\ref{fself}), in a more useful manner.  Focusing on the constant
piece of $\beta(\Delta)$, we write
\begin{equation}
\Sigma_{\mu\nu}=-M^2\Pi_{\mu\nu}=-M^2\left(
g_{\mu\nu}+\nabla_\mu\fft1\Delta\nabla_\nu\right),
\end{equation}
and assume that it acts between vector quantities, $\phi^\mu(x)
\Sigma_{\mu\nu}(x,y)\phi^\nu(y)$.  Since this would be integrated
over all of space in order to compute physical quantities, we may
integrate by parts to obtain
\begin{eqnarray}
\int d^4x\,d^4y\,\phi^\mu(x)\Sigma_{\mu\nu}(x,y)\phi^\nu(y)&=&\\
&&\kern-6em
-M^2\int d^4x\,d^4y\,\phi^\mu(x)\left(g_{\mu\nu}\delta^4(x-y)-(\nabla_{x^\mu}
\nabla_{y^\nu}\Delta^{-1}(x,y))\right)\phi^\nu(y),\nonumber
\end{eqnarray}
where $\Delta^{-1}$ is the $E_0=3$ (minimally coupled) scalar Greens
function in an AdS background.

Working in homogeneous coordinates, where the scalar Greens function
has the form
\begin{equation}
\Delta^{-1}(Z)=\fft1{4\pi^2L^2}\left(\fft{Z}{Z^2-1}+\fft12\log\fft{Z+1}{Z-1}
\right)\qquad(E_0=3),
\end{equation}
and dropping the contact term, we find that the non-local part of
$\Sigma_{MN}$ has the form
\begin{equation}
\Sigma_{MN}=-M^2\nabla_M\nabla_N\Delta^{-1}(Z)=-
\frac{M^2}{2\pi^2 L^4(Z^2-1)^2}\left(\hat{G}_{MN}-4 Z N_M N_N \right).
\end{equation}
This may be expanded in terms of the asymptotic bi-vectors given in
Table~\ref{tbl:1}.  The result is
\begin{equation}
\Sigma_{MN}=-\frac{M^2}{2\pi^2 L^4}\left(3T_3+6T_5+9T_7 \ldots\right).
\label{eq:mextract}
\end{equation}
Note that only odd $T_i$'s show up in this expression.  Finally, we see
that the mass may be extracted from the leading asymptotic behavior of 
the self-energy according to $\Sigma_{MN}\sim (-3M^2/2\pi^2L^4)T_3+
\cdots$.  In the next section we will compute scalar and spin-1/2 loop
contributions to $\Sigma$, and in this manner find explicit expressions
for the induced vector mass.


\section{A one-loop computation of the graviphoton mass}

For a non-abelian vector with gauge group $G$ coupled to a conserved
current $J^a_M(X)$, the self-energy is given by the two-point function
$\Sigma_{MN}^{ab}(X,Y) =\langle J_M^a(X)J_N^b(Y)\rangle$.  Here $a$ and
$b$ denote gauge indices taking values in the adjoint of $G$.  We take
the currents to have the forms
\begin{eqnarray}
J^a_{(0)\,M}&=&\frac{ig}{2}\phi_i\left(\overrightarrow{\nabla}_M-
               \overleftarrow{\nabla}_M\right)\left(T^a\right)_{ij}\phi_j,
\nonumber\\
J^a_{(\fft12)\,M}&=&ig\overline{\psi_i}\Gamma_M
               \left(S^a\right)_{ij}\psi_j ,
\end{eqnarray}
for couplings to real spin-0 and Dirac spin-1/2 fields, respectively.
We take $T^a$ and $S^a$ to generate arbitrary (potentially reducible)
representations of $G$, except that $T^a$ must generate only real
representations.

Evaluating the self-energy follows from Wick's theorem.  In order to
proceed, we need explicit forms for the scalar and spinor propagators
in AdS.  Since in the end we are interested in conformal matter, we
consider only conformally coupled scalars and massless Dirac fields.
Their respective propagators are given by \cite{Avis:1977yn}
\begin{eqnarray}
\label{prop0}
\Delta_{(0)}&=&\frac{1}{8\pi^2
  L^2}\left(\frac{\alpha_+}{Z+1}+\frac{\alpha_-}{Z-1} \right),\\
\label{prop.5}
\Delta_{(\frac{1}{2})}&=&\frac{1}{8\pi^2L^4}\left(\frac{\alpha_+
\Gamma^M(X_M-Y_M)}{(Z+1)^2}+\frac{\alpha_-\Gamma^M(X_M+Y_M)}{(Z-1)^2} \right),
\end{eqnarray}
where $\alpha_+$ and $\alpha_-$ specify boundary conditions on AdS$_4$
\cite{Porrati:2001db,Duff:2002ab}.  At short distances compared to the
AdS radius $L$, the behavior of these propagators must match onto the
corresponding flat-space ones.  Since short distances correspond to
$Z\to-1$, and the properly normalized four-dimensional behavior
must be of the form
\begin{equation}
  \Delta(Z\to-1)\sim-\fft{1}{4\pi^2|X-Y|^2}=\frac{1}{8\pi^2 L^2}\frac{1}{Z+1},
\end{equation}
we see that this demands $\alpha_+=1$.  On the other hand, $\alpha_-$ is
undetermined at short distances.  The expressions (\ref{prop0}) and
(\ref{prop.5}) clearly indicate an image charge structure on the Einstein
static universe, with $\alpha_-=\pm1$ corresponding to reflecting boundary
conditions and $\alpha=0$ to transparent ones.

We again concern ourselves only with the non local part, and so we only
take Wick contractions between fields at different points.  We take the
propagators to be diagonal in representation space, so that Wick contractions
give $\delta_{ij}$ in addition to the propagators of (\ref{prop0}) and
(\ref{prop.5}).  For the spin-0 contribution to the self-energy, we find
\begin{eqnarray}
\hat{\Sigma}^{ab}_{(0)\;MN}
    &=&-\ft14g^2
       \left<\phi_i\nabla_M\phi_j\phi_k\nabla_N\phi_l\right>
       \big((T^a)_{ij}-(T^a)_{ji}\big)\big((T^b)_{kl}-(T^b)_{lk}\big) 
                   \nonumber \\
    &=&g^2 \Tr(T^a T^b) \left(-\Delta_{(0)}\nabla_M\nabla_N\Delta_{(0)}+
      \nabla_M\Delta_{(0)}\nabla_N\Delta_{(0)}\right).
\end{eqnarray}
Working out the derivatives of the scalar propagator, (\ref{prop0}), and
dropping contact terms (since we are only interested in the long distance
behavior), we find
\begin{eqnarray}
\label{eq:M1.0exp}
\hat{\Sigma}^{ab}_{(0)\;MN}
    &=&\frac{-g^2\Tr(T^aT^b)}{64\pi^4L^6}
        \Bigg[
        \frac{\alpha_{+}^2}{(Z+1)^3}\left(\left(\hat{G}_{MN}-ZN_M
        N_N\right)+ N_M N_N\right)  \nonumber \\
      & &\kern2.5cm +\frac{\alpha_{-}^2}{(Z-1)^3}\left(\left(\hat{G}_{MN}-ZN_M
        N_N\right)- N_M N_N\right) \nonumber \\
      & &\kern2.5cm +\frac{2\alpha_{+}\alpha_{-}Z}{(Z^2-1)^2}
         \left(\left(\hat{G}_{MN}-ZN_M N_N\right)-3N_M N_N\right)\Bigg]
\nonumber\\
      &=&\frac{-g^2\Tr(T^aT^b)}{64\pi^4L^6}\left[
                 \left((\alpha_{+}^2-\alpha_{-}^2)T_3- 
                 3(\alpha_{+}^2+\alpha_{-}^2)T_4\ldots\right)
           +2\alpha_{+}\alpha_{-}\left(-3T_4\ldots\right)\right].\quad
\end{eqnarray}
Since the induced photon mass is proportional to the coefficient of
$T_3$, this demonstrates that it must vanish when reflecting boundary
conditions are imposed on the scalars, $\alpha_+^2=\alpha_-^2$.  A
mass is generated for all other cases (which incidentally correspond to
non-unitary behavior at the field theory level, since in such cases not
all of the information is reflected back from the AdS$_4$ boundary).

A similar computation for the fermion-loop contribution yields
\begin{eqnarray}
\hat{\Sigma}^{ab}_{(\frac{1}{2})\;MN}
      &=&-g^2\Tr(S^aS^b)\Tr
                             \left(\Gamma_M\Delta_{(\frac{1}{2})}(X,Y)
                             \Gamma_N\Delta_{(\frac{1}{2})}(Y,X)
                              \right)\nonumber \\ 
      &=&\frac{-g^2\Tr(S^aS^b)}{8\pi^4 L^6}
        \Bigg[
        \frac{\alpha_{+}^2}{(Z+1)^3}\left(\left(\hat{G}_{MN}-ZN_M
        N_N\right)+ N_M N_N\right)  \nonumber \\
        & &\kern2.5cm
+\frac{\alpha_{-}^2}{(Z-1)^3}\left(\left(\hat{G}_{MN}-ZN_M
        N_N\right)- N_M N_N\right)\Bigg]\nonumber \\
      &=&
       \frac{-g^2\Tr(S^aS^b)}
                  {8\pi^4 L^6}\left[(\alpha_{+}^2-\alpha_{-}^2)T_3- 
                     3(\alpha_{+}^2+\alpha_{-}^2)T_4\ldots\right].
\label{eq:M1.5exp}
\end{eqnarray}
This expression is similar to that for the scalar loop, except for the
absence of a mixed $\alpha_+\alpha_-$ term.  This appears to be a result
of working with Dirac spinors, and may not hold for Majorana ones.
Nevertheless, this potential mixed term is irrelevant as far as mass
generation is concerned, since it is of higher order in the asymptotic
expansion.

As a result, we may obtain a universal expression for the induced graviphoton
mass based on its coupling to conformal spin-0 and spin-1/2 fields.  In
order to preserve supersymmetry, the boundary conditions of all fields in
a single supermultiplet must be chosen identically \cite{Breitenlohner:jf}.  
This allows us to combine the results of (\ref{eq:M1.0exp}) and
(\ref{eq:M1.5exp}).  Comparing with the leading behavior of (\ref{eq:mextract}),
we therefore find that the dynamically generated mass of the graviphoton
is given by
\begin{equation}
(M_1)^2=g^2\frac{(\alpha_{+}^2-\alpha_{-}^2)}{96\pi^2L^2}\left(
\Tr(T^aT^b)+4\Tr(S^aS^b)\right),
\label{eq:m1mass}
\end{equation}
where the traces are now taken over real scalars (representation $T^a$)
and Majorana fermions (representation $S^a$), respectively.


\section{A check of supersymmetry}

In general, unitary representations of AdS$_4$, given by $D(E_0,s)$, must
satisfy a bound $E_0 \geq s+1$ for $s \geq 1$.  Saturation of this
bound corresponds to shortened (`massless') representations.  As indicated
in (\ref{eq:adshlim}), the AdS Higgs mechanism is related to the fact
that a massive representation becomes reducible in the limit
$E_0 \rightarrow s+1$.

There is of course a supersymmetric generalization of the AdS Higgs
phenomenon, where a complete supermultiplet may pick up a mass without
explicitly breaking supersymmetry.  Note that supersymmetry relates
$E_0$ and $s$ by units of $1/2$ within a supermultiplet.  In particular,
the massless graviton and graviphotons are given by $D(3,2)$ and $D(2,1)$,
respectively.  While the massive representations of supersymmetry tend to
be rather large, we will not need any explicit forms.  We simply note
that, for the supergravity multiplet including graviphotons, the AdS Higgs
phenomenon yields the partial decomposition
\begin{equation}
D(3+\epsilon,2)+D(2+\epsilon,1)+\cdots
=(D(3,2)+D(4,1))+(D(2,1)+D(3,0))+\cdots,
\end{equation}
in the limit that $\epsilon \rightarrow 0$.  The ellipses on both
sides denote other members of the supermultiplet that we are not explicitly
interested in.  Note, however, that the mass shift (or rather $E_0$ shift)
is given by the {\it same} parameter $\epsilon$ for all members of the
multiplet.

In this manner, we may check whether the graviphoton mass computed here
agrees with the graviton mass computed previously in
\cite{Porrati:2001db,Duff:2002ab}.  Using the relations between mass
and $E_0$ \cite{Englert:1983rn}
\begin{equation}
E_0^{(s = 1)} = \ft32+\ft12\sqrt{1+(M_1L)^2},\qquad
E_0^{(s = 2)} = \ft32+\ft12\sqrt{9+(M_2L)^2},
\end{equation}
and expanding for small masses, we find
\begin{equation}
\epsilon^{(s=1)}=\ft14(M_1L)^2,\qquad
\epsilon^{(s=2)}=\ft1{12}(M_2L)^2.
\end{equation}
If $\epsilon$ were universal, as demanded by supersymmetry, this would
indicate that $(M_2)^2=3(M_1)^2$.  We now recall the result of
\cite{Duff:2002ab}, which gives
\begin{equation}
\label{eq:M2calc}
(M_2)^2=8\pi G_4 \frac{(\alpha_{+}^2-\alpha_{-}^2)}{160\pi^2L^4}
                 (n_0+3n_{\frac{1}{2}}+12n_1),
\end{equation}
for the mass of the graviton.  Here $G_4$ is the four-dimensional Newton's
constant, and $n_0$, $n_{\fft12}$ and $n_1$ are the total number of real
spin-0, Majorana spin-1/2 and spin-1 fields.  It is interesting to note
that the cross terms proportional to $\alpha_{+}\alpha_{-}$ vanish in the
graviton mass calculation because there is a relative sign difference
in the $\alpha_{-}$ boundary condition between scalars and pseudoscalars,
and they come in equal numbers in a supermultiplet.  On the other hand,
while such cross terms are present in the graviphoton calculation, they
simply do not contribute to the mass term.

Finally, we combine (\ref{eq:m1mass}) with (\ref{eq:M2calc}) and the
supersymmetry condition $(M_2)^2=3(M_1)^2$ to obtain 
\begin{equation}
\label{eq:condition}
\fft{8\pi G_4}{5L^2}(n_0+3n_{\fft12}+12n_1)
=g^2(S_2(T_0)+4S_2(S_{\fft12})),
\end{equation}
which is the resulting condition for unbroken supersymmetry.  We have
defined $\Tr(R^aR^b)=S_2(R)\delta^{ab}$, which is essentially
the sum of the indices of the irreducible representations comprising $R$.
While the number of fields, $n_0$ and $n_{\fft12}$, do not show up
explicitly on the right hand side, this information is contained in the
fact that the scalar and spinor representations may be reducible.
In particular, $n_0$ and $n_{\fft12}$ simply counts
the dimensions of the representations, so that $n_0={\rm dim}\,(T_0)$
and $n_{\fft12}={\rm dim}\,(S_{\fft12})$.  We have not considered vectors
in the loop, as consistency of the non-abelian theory demands a uniform
treatment of such vectors, and we do not modify the reflecting boundary
conditions of the graviphotons themselves (so that there is no spin-1
contribution to the mass on the right hand side).

We now look at some specific examples of gauged supergravities admitting
an AdS$_4$ vacuum solution.  For matter coupled supergravities, there is
often a choice of how much gauging may be turned on.  Perhaps the simplest
cases involve gauging only the $R$ symmetry.  However, even here one may
not have enough graviphotons to gauge the entire group, so that only a
subgroup of the full $R$ symmetry may be gauged.  After gauging, the
graviphotons transform under the adjoint of the gauged $R$ symmetry group,
and couple to the corresponding $R$ symmetry currents of the matter
sector.  In addition, turning on the gauging leads to a potential; for
the theories of interest, the potential admits a stable AdS$_4$ vacuum.
Since the strength of the potential is related to the gauge coupling,
this allows us to rewrite $g^2$ in terms of the AdS$_4$ radius $L$ and
the four-dimensional Newton's constant $G_4$.  Noting that $g^2\sim
G_4/L^2$, this allows a direct comparison of both sides of
Eq.~(\ref{eq:condition}).

In the context of large $N$ gauge theories coupled to supergravity, it is
only the matter fields that are expected to receive unusual ({\it i.e.}
transparent) boundary conditions.  As a result, only matter in the loop
would give rise to dynamically generated masses, and we may ignore gravity
multiplet self-contributions to its mass.  In any case, potential graviton
loop effects would be suppressed by $1/N^2$, and at least formally may be
ignored.

The simplest example of AdS$_4$ gauged supergravity is the ${\cal N}=2$
model, where the $O(2)$ symmetry may be gauged by the graviphoton of the
${\cal N}=2$ gravity multiplet \cite{Fradkin:1976xz,Freedman:1976aw}.
Following the convention of \cite{Luciani:1977hp}, which couples this
model to an arbitrary number of vector multiplets, we see that
gauging yields a constant negative potential $-3g^2/2K^4$ where
$4K^2=16\pi G_4$.  Relating this potential to the cosmological constant,
we find the gauge coupling constant to be given by $g^2=4\pi G_4/L^2$,
so that the condition (\ref{eq:condition}) becomes
$\fft25(n_0+3n_{\fft12}+12n_1)=(S_2(T_0)+4S_2(S_{\fft12}))$.  The
${\cal N}=2$ vector multiplet is composed of one vector field, 2 Majorana
spinors, and 2 scalars (one is a pseudoscalar).  However only the
Majorana spinors are charged under $O(2)$, transforming with
generators $i\epsilon^{lm}$.  This results in $S_2(S_{\fft12})=2$
(as we should expect, since this Majorana doublet under $O(2)$ corresponds
to one Dirac spinor under $U(1)$).  With this counting, we find that
the supersymmetry condition is met.  Note, however, that while all
members of the vector multiplet contribute to the graviton mass, only
the gauginos contribute to the graviphoton mass.

Turning next to ${\cal N}=4$ gauged supergravity, we recall that there
are two versions of the ungauged theory, with global $SO(4)$ and $SU(4)$ 
symmetries, respectively.  Gauging of the latter theory yields the
$SU(2)\times SU(2)$ Freedman-Schwarz model \cite{Freedman:ra} with no
stable extrema.  On the other hand, gauging the $SO(4)\simeq
SU(2)\times SU(2)$ model using all six graviphotons yields a potential
admitting an AdS$_4$ solution \cite{Das:pu,Gates:1982ct}.  It was further
shown in \cite{Cvetic:1999au} that different gauge couplings in each of
the $SU(2)$'s may be obtained by field redefinitions of the theory where
the gauge couplings were initially taken to be the same.  Therefore,
without loss of generality, we set the two $SU(2)$ couplings to be equal.

For the standard ${\cal N}=4$ gauged model coupled to vector multiplets,
the resulting potential receives contributions from the scalars in the
vector multiplet.  However, the AdS$_4$ extremum is given by the constant
part of the potential, $V_0= -12g^2/(16\pi G_4)$.  Rewriting this
in terms of the AdS$_4$ radius yields $g^2=\frac{16\pi G_4}{2L^2}$, and
a resulting supersymmetry condition $\fft15(n_0+3n_{\fft12}+12n_1)
=(S_2(T_0)+4S_2(S_{\fft12}))$.  Before proceeding, we must be careful to
state how each field transforms under the $SU(2)\times SU(2)$ gauge group.
The spinors transform as a $(2,2)$ and the scalars as a $(1,3)+(3,1)$.  We
use $i\frac{\overrightarrow{\sigma}}{2}$ and $i\epsilon^{ijk}$ as the
generators of $SU(2)$ working on spinors and scalars respectively.  This
gives that $S_2(T_0)=2$ and $S_2(S_{\fft12})=1$ (for either one of the two
$SU(2)$ factors).  The check of supersymmetry then reads
\begin{equation}
\ft15(6+3\times 4+12\times 1)=(2+4\times 1),
\end{equation}
which is indeed satisfied.  As in the ${\cal N}=2$ case above, the
contributions to the graviton and graviphoton masses arise in different
combinations.  However, the factors for the complete vector multiplet
conspire to ensure supersymmetric mass generation for the gravity multiplet.


\section{Discussion}

Although much of the investigation on the AdS Higgs phenomenon has been
motivated by holography of the Karch-Randall model, it is worth emphasizing
that the phenomenon does not depend on extra dimensions or other novel
ideas, and is simply a result of looking at field theory in curved spacetimes.
The reason the Higgs phenomenon has gone unnoticed for a long time is
because it only shows up in the presence of unusual (and typically
non-unitary) boundary conditions on the fields in AdS$_4$, despite the
fact that transparent boundary conditions were known since at least the
work of \cite{Avis:1977yn}.

Of course, non-unitary boundary conditions are often dismissed as
unphysical.  However, another interpretation may be given: that the
theory in AdS$_4$ by itself is incomplete, and that one must also
include a defect field theory on the boundary of the AdS$_4$ spacetime
\cite{DeWolfe:2001pq,Erdmenger:2002ex}.  This point of view is natural
in terms of holography with multiple domains, such as what happens in
the Karch-Randall model \cite{Porrati:2001gx,Bousso:2001cf}.  In this
context, the condition for unbroken supersymmetry that we have derived,
(\ref{eq:condition}), provides an important check when considering the
viability of a supersymmetric theory living on the brane.  In particular,
it suggests that the construction of an ${\cal N}=4$ theory living on the
Karch-Randall brane is indeed possible.

This dynamically generated mass for the supergravity multiplet scales
as $G_4/L^4$, and vanishes in the flat space limit of AdS$_4$.  This is
of course to be expected, because in this limit, the focusing effect
of anti-de Sitter space is lost, and the composite Goldstone boson
responsible for the Higgs mechanism is no longer kinematically bound
together.  At the same time, boundary conditions at infinity lose their
importance when AdS$_4$ is reduced to flat space.

We note that, while we have not directly computed the one-loop
gravitino mass, such a calculation would be straightforward and may be
related to the two-point function of the supercurrent,
$\langle J_\mu^\alpha(x) J_\nu^\beta(y)\rangle$.  Since the supercurrent
is in the same multiplet as the stress tensor and $R$-current, we expect
the resulting gravitino mass to also respect the supersymmetry condition
discussed in the previous section.  Although it may be somewhat surprising
to consider unbroken supersymmetry with massive gravitinos, we emphasize
that this is no more so than unbroken general covariance with a massive
graviton or unbroken gauge invariance with massive photons.

Finally, we emphasize that the AdS Higgs phenomenon consistently
generates masses for non-abelian gauge fields, and not just for
abelian photons.  This is clear from the example of gauged ${\cal N}=4$
supergravity, where the $SU(2)\times SU(2)$ graviphotons become massive.
Although this Higgs phenomenon is purely an anti-de Sitter effect, related
to unusual boundary conditions, it nevertheless provides a new gauge
invariant mechanism for mass generation for non-abelian gauge bosons.  Of
course, such masses are generally small, on the order of the natural AdS
scale.  However, they may be controlled by making adjustments to the matter
fields and their boundary conditions.  While an actual AdS spacetime is
disfavored by observation, it would still be curious to see if this Higgs
phenomenon has any relevance to mass generation in the Standard Model.

\bigskip
\acknowledgments

This work was supported in part by the US Department of Energy under
grant DE-FG02-95ER40899.  JTL wishes to thank M.~Duff, M.~Porrati and
H.~Sati for discussions and D.~Gross for raising the issue of mass
generation for photons in AdS and for highlighting the kinematically
bound state nature of the Goldstone boson.


\end{document}